\documentclass[12pt]{iopart}
\usepackage[dvips]{epsfig}
\usepackage[usenames]{color}
\usepackage{pstricks}
\usepackage{amssymb}
\usepackage{subfigure}
\usepackage{afterpage}
\usepackage[utf8]{inputenc}
\usepackage{bm}

\begin{document}

\title{Bound and scattering states in harmonic waveguides in the vicinity of free space Feshbach resonances}

\author{Gaoren Wang$^1$, Panagiotis Giannakeas$^2$, Peter Schmelcher$^{1,3}$}

\address{$^1$ Zentrum für Optische Quantentechnologien, Universität Hamburg, Luruper Chaussee 149, 22761 Hamburg, Germany}

\address{$^2$ Department of Physics and Astronomy, Purdue University, West Lafayette, Indiana 47907, USA}

\address{$^3$ The Hamburg Centre for Ultrafast Imaging, Universität Hamburg, Lurper Chaussee 149, 22761 Hamburg, Germany}
\vspace{10pt}

\begin{abstract}
The two-body bound and scattering properties in an one-dimensional harmonic 
waveguide close to free space magnetic Feshbach resonances are investigated based 
on the local frame transformation approach within a single partial wave approximation. 
An energy and magnetic field 
dependent free space phase shift is adopted in the current theoretical framework. 
For both $s$- and $p$-wave interaction, the least bound state in the waveguide 
dissociates into the continuum at the resonant magnetic field where 
the effective one-dimensional scattering length $a_{\rm 1D}$ diverges.
Consequently, the association of atoms into molecules in the waveguide  
occurs when the magnetic field is swept adiabatically across the pole of $a_{\rm 1D}$. 
In the vicinity of broad $s$-wave resonances, the resonant magnetic field 
is nearly independent on the transverse confining frequency $\omega_{\perp}$ of the waveguide. 
Close to $p$-wave and narrow $s$-wave resonances, the resonant magnetic field 
changes as $\omega_{\perp}$ varies.
\end{abstract}

%
%
%
%
%

\section{Introduction}
Ultracold atoms in deep optical lattices offer a highly controllable platform to
study quantum systems in reduced dimensions \cite{rmp:80:885}.
The shape of the optical lattices can be engineered to realize different confining geometries \cite{rpp:76:086401} 
and the interatomic interaction can be tuned through 
magnetic Feshbach resonances \cite{rmp:82:1225}.
As a result, ultracold atoms in quasi-one dimensional (quasi-1D) waveguides have been
realized experimentally \cite{prl:94:210401,prl:95:230401,prl:104:153203,prl:110:203202}  
where the reduced dimensionality strongly affects the two-body bound and 
scattering states. In the waveguide, weakly bound molecular states exist for both positive 
and negative scattering length \cite{prl:94:210401}. This is in contrast to free space where the weakly bound 
states exist only on the side of positive scattering length \cite{rmp:82:1225}. 
A prominent scattering property in waveguides is the confinement-induced resonance (CIR) 
\cite{prl:104:153203, prl:81:938, prl:91:163201} which is characterized by 
the divergence of the effective 1D interaction strength $g_{\rm 1D}$, and occurs at 
finite free space scattering length \cite{JPIV:116:69}.\\
\indent Theoretically, the $s$-wave \cite{pra:71:012709, pra:72:053625, pra:73:052709, pra:86:033601, pra:91:042703} 
and $p$-wave \cite{pra:91:043622} bound states in the waveguide close to 
Feshbach resonances have been studied based on two-channel model. The resonant 
scattering properties in the waveguide have been investigated numerically by using 
a multichannel model \cite{pra:86:062713}. In this work, the 
local frame transformation (LFT) \cite{pra:24:619,prl:49:128,pra:36:4236} will be utilized to 
explore the two-body collisions in a harmonic waveguide. It has been shown 
\cite{prl:92:133202, pra:86:042703, prl:111:183201, pra:88:012715, pra:89:052716, pra:92:022706} 
that the LFT method is convenient to tackle high partial wave collisions in confined geometries 
and also to deal with couplings of different partial wave states due to the confinement.
In previous studies using the LFT method \cite{pra:89:052716, pra:92:022706}, the interatomic interaction is 
described by a single-channel potential. The short-range part of the potential 
is modified to mimick the variation of the phase shift close to Feshbach resonances. 
Within such a treatment, the influence of the resonance width is not accounted for.
In the current work, the energy and magnetic field dependent phase shift induced by the 
multichannel interatomic interaction is incorporated into the LFT method, 
which allows one to calculate the two-body properties in the waveguide close to both broad 
and narrow resonances. The current LFT approach complements previous methods 
\cite{pra:72:053625, pra:73:052709, pra:86:033601, pra:91:042703} which have been dedicated to 
explore the confinement induced Feshbach molecules.\\
\indent The weakly bound molecular state in a harmonic waveguide can be tuned by 
a magnetic field, and crosses the scattering threshold at the magnetic field 
where the effective 1D scattering length $a_{\rm 1D}$ diverges. Accordingly, 
in a harmonic waveguide, the association of an unbound atom pair into a molecule 
occurs when the magnetic field is swept adiabatically across the pole of $a_{\rm 1D}$. 
For $s$-wave 
collisions, $g_{\rm 1D}$ is inversely proportional to $a_{\rm 1D}$ \cite{oc:243:3}. 
Hence, the molecular association occurs at $g_{\rm 1D}=0$ other than at CIR. 
For $p$-wave collisions, $g_{\rm 1D}$ 
is proportional to $a_{\rm 1D}$ \cite{oc:243:3}, and the molecular association occurs at the CIR. 
Here we assume that the two atoms possess the same mass. If the two masses 
are different, the center of mass and relative motions are 
coupled and the corresponding molecular formation process has been discussed 
in \cite{njp:11:073031}.\\
\indent Our work is organized as follows. In Sec.~II, we briefly review 
the local frame transformation approach, and show how the two-body properties 
in a waveguide are derived. 
Sec.~III introduces the free space phase shift which is an ingredient of the LFT approach. 
In Sec.~IV, the bound state and scattering properties 
in the waveguide are presented and discussed.  
Sec.~V provides our conclusions.
\section{The local frame transformation approach}

\indent We consider ultracold collisions of two identical atoms in a harmonic waveguide with cylindrical symmetry. 
An external magnetic field is applied to tune the interatomic interaction via free space Feshbach resonances \cite{rmp:82:1225}. 
The direction of the magnetic field is assumed to be parallel to the symmetry axis of the waveguide, 
namely the $z$ axis. 
Due to the harmonic confinement, the center of mass motion and relative motion are 
separable. Due to this separation, all the relevant collisional physics is described 
by the Hamiltonian of the relative degrees of freedom which reads as follows
\begin{eqnarray}
H=-\frac{\hbar^2}{2\mu}\nabla^2+\frac{1}{2}\mu\omega_{\perp}^2\rho^2+V(\bm{r}),
\end{eqnarray}
where $\mu$ is the reduced mass, $\omega_{\perp}$ is the transverse confinement frequency, 
$r=\sqrt{z^2+\rho^2}$ is the interatomic distance with $z$ and $\rho$ being the 
longitudinal and transverse components of the vector 
$\bm{r}$, respectively. $V(\bm{r})$ is the interatomic interaction.\\
\indent The local frame transformation \cite{pra:89:052716, pra:92:022706} is employed to study
the two-body properties in the waveguide. The key property of the LFT relies on 
the length scale separation between the confining potential and the interatomic interation potential. 
More specifically, the length scale of the confinement is characterized by the harmonic
length $a_{\perp}=\sqrt{\hbar/\mu\omega_{\perp}}$.
The long-range interactomic interaction is considered to be the isotropic 
van der Waals interaction, and the anisotropic interaction such as 
magnetic dipole-dipole interaction is neglected.
Then the length scale of the interatomic interaction is given by $\beta_6=(2{\mu}C_6/\hbar^2)^{1/4}$, 
where $C_6$ is van der Waals coefficient  \cite{rmp:82:1225}.
The validity of the LFT is ensured by the length scale separation $a_{\perp}\gg{\beta_6}$.
Due to this separation, two regions
with different symmetries are identified. In the region $r\sim\beta_6$, the
interatomic interaction is dominant and the confinement potential is negligible.
The two-body collision is treated as free space scattering problem at energy $E$. The spherical
symmetry in this region is exploited as the wavefunction is expanded in partial wave states. 
A single partial wave approximation is adopted which is justified due to the low energy collison considered here.
All the scattering information in this region is contained in the free space $K$ matrix $K^{3D}=\tan\eta$, 
where $\eta$ is the free space scattering phase shift.
To make sure that the phase shift $\eta$ is well defined, the total energy $E$
is considered to be positive $E>0$. 
In the region $r{\ge}a_{\perp}$, the atoms feel the confinement potential whereas the
interatomic interaction is negligible. The system possesses cylindrical symmetry, 
and the corresponding wavefunction is expressed as the product of the harmonic oscillator wavefunction
in the transverse plane and a 1D plane wave in the
longitudinal $z$-direction. 
The $K$ matrix in this region is denoted as $K^{1D}$, which contains the bound and scattering properties in the presence
of the waveguide. In the intermediate region $\beta_6<r<a_{\perp}$,
the local frame transformation matrix $U$ \cite{pra:24:619, prl:49:128, pra:36:4236} is used to connect the wavefunctions 
at short distances to that at large distances. 
The $K^{1D}$ matrix in the waveguide is related to the free space $K^{3D}$ matrix by $K^{1D}=UK^{3D}U^T$ \cite{pra:89:052716}. 
The elements of the local transformation matrix read \cite{pra:89:052716}
\begin{equation}
U^T_{l, n}=\frac{\sqrt{2}(-1)^{d_0}}{a_{\perp}}\sqrt{\frac{2l+1}{kq_n}}P_l\left(\frac{q_n}{k}\right),
\label{eq_LocalTransformationMatrixElement}
\end{equation}
where $l$ is the partial wave quantum number, and $n$ is the quantum number
specifying the harmonic oscillator state in the transverse plane. 
$d_0$ equals $l/2$ when $l$ is even, and is equal to $(l+1)/2$ when $l$ is odd. 
$P_l(x)$ is the $l$th-order Legendre polynomial, and 
$q_n$ is the longitudinal momentum given by 
$\frac{(\hbar{q_n})^2}{2\mu}=E-\hbar\omega_{\perp}(2n+1)$. 
The quantum number related to the azimuthal symmetry in transverse waveguide directions
is set to zero in the derivation of equation~(\ref{eq_LocalTransformationMatrixElement}) \cite{pra:89:052716}.\\
\indent For the energy region $0<E<\hbar\omega_{\perp}$, i.e.\ the total
energy $E$ is smaller than the ground state energy of the transverse confinement,
a possible bound state in the waveguide is determined by
\begin{equation}
\det(I-iK^{1D})=0.
\label{eq_boundlevelinwaveguide}
\end{equation}
For the energy region $\hbar\omega_{\perp}<E<3\hbar\omega_{\perp}$ in which
only the ground mode of the transverse confinement is occupied, the
physical $K^{\rm 1D, phys}$ matrix is defined by properly eliminating all
the energetically closed excited modes \cite{pra:89:052716}
\begin{equation}
K^{1D,\rm phys}=\frac{{K^{3D}}U_{l,n=0}^2}{1-iK^{3D}\mathfrak{U}_{l,l}},
\label{eq_K1DPhys}
\end{equation}
where $\mathfrak{U}_{l,l}=\sum\limits_{n=1}^{\infty}U_{l,n}^2$. \\
\indent The collision in the waveguide can be mapped to a one-dimensional
scattering problem \cite{JPIV:116:69}.
For $s$-wave collisions in the waveguide, the effective 1D scattering length $a_{\rm 1D}^s$
defined in \cite{prl:81:938} is related to 
$K^{\rm 1D, phys}$ by 
\begin{equation}
a_{\rm 1D}^s=\lim_{q_0\rightarrow{0}}1/(q_0K^{\rm 1D, phys}). 
\label{eq_a1Ds}
\end{equation}
The 1D interaction strength $g_{\rm 1D}^s$ is expressed in terms of $a_{\rm 1D}^s$ by
$g_{\rm 1D}^s=-\hbar^2/(\mu{a_{\rm 1D}^s})$ \cite{JPIV:116:69}. 
According to the definition in \cite{prl:100:170404},
the effective 1D scattering length for $p$-wave collisions in the waveguide is expressed as 
\begin{eqnarray}
a_{\rm 1D}^p=-\lim_{q_0\rightarrow{0}}K^{\rm 1D, phys}/q_0. 
\label{eq_a1Dp}
\end{eqnarray}
In contrast to the $s$-wave case in which $g_{\rm 1D}^s$ is inversely proportional to 
$a_{\rm 1D}^s$, the effective 1D interaction strength for $p$-wave collisions 
is $g_{\rm 1D}^p=-\hbar^2a_{\rm 1D}^p/\mu$ i.e.\ is proportional to $a_{\rm 1D}^p$ \cite{oc:243:3}.

\section{Free space phase shift}
\indent As shown by equations~(\ref{eq_boundlevelinwaveguide})-(\ref{eq_a1Dp}), 
the free space $K^{\rm 3D}$ matrix or equivalently the free space phase shift is needed to calculate the bound and 
scattering properties in the waveguide using the LFT method. In the ultracold regime, 
the energy and magnetic field dependent phase shift in the vicinity of a well-isolated 
$s$-wave Feshbach resonance is obtained via \cite{prl:108:250401,prl:108:045304}
\begin{equation}
-\frac{\tan\eta_s(E,B)}{k}=a_{\rm bg}\left[1+\frac{\Delta}{E/\delta\mu-(B-B_0)}\right],
\label{eq_sWavePhaseShift}
\end{equation}
where $\hbar{k}=\sqrt{2\mu{E}}$, $B$ is the magnetic field strength,  
and $a_{\rm bg}$ is the background scattering length. 
$B_0$ and $\Delta$ are the resonance position and width, respectively. 
$\delta\mu$ is the difference between the magnetic moments of the incident 
scattering state and the bound state which produces the resonance. 
The two-body properties in the waveguide around two $^6$Li $s$-wave 
Feshbach resonances are explored in the following. The values of $B_0$, $\Delta$,
$\delta\mu$, $a_{\rm bg}$ and $C_6$ for the two $s$-wave resonances are listed in
Table \ref{table:parameter_FeshbachResonance}. The parameter $R^{*}=\frac{\hbar^2}{2{\mu}a_{\rm bg}\delta\mu\Delta}$ \cite{prl:93:143201}
is also given in the table. $R^{*}\ll1$ indicates a broad resonance, 
and the closed-channel component in the wavefunction is small over the 
resonant width. $R^{*}\gg1$ indicates a narrow resonance, 
and the closed-channel component in the wavefunction is dominant over a 
large fraction of the resonant width. 
Hereafter, the broad $^6$Li Feshbach resonance at
$B_0=832$~G will be referred to as resonance I, and the narrow $^6$Li Feshbach resonance
at $B_0=534.4$~G is denoted as resonance II. \\
\indent The collision of two $^{40}$K atoms in the waveguide close to a $p$-wave Feshbach resonance \cite{pra:69:042712} 
is also studied in Section IV. The energy and magnetic field-dependent phase shift obtained in \cite{pra:69:042712} 
is utilized in the calculation of the bound and scattering properties. 
\begin{table}[]
\caption{Parameters of the two $^6$Li Feshbach resonances.
The data are taken from \cite{pra:89:042701}.
}
\centering
\begin{tabular}{ l c c c c c c}
\hline\hline\\
				&		$B_{0}$ [G]	&	$\Delta$ [G]	&	$\delta\mu$ [$\mu_B$]	&	$a_{\rm bg}$ [au]	&	$C_6$ [au]	&	$R^{*}$ [au]		\\
\hline \\
Resonance I		&		832					&	-262			&	1.87					&	-1593				&	1393.39		&	 0.55				\\
Resonance II	&		534.4				&	0.1				&	1.97					&	59					&	1393.39		&	 3.69$\times10^4$    \\
\hline\hline
\end{tabular}
\label{table:parameter_FeshbachResonance}
\end{table}

\section{Bound and scattering properties in waveguides}
\subsection{$s$-wave case}
\begin{figure}
\centering
\begin{minipage}[b]{0.5\textwidth}
\includegraphics[width=1\textwidth]{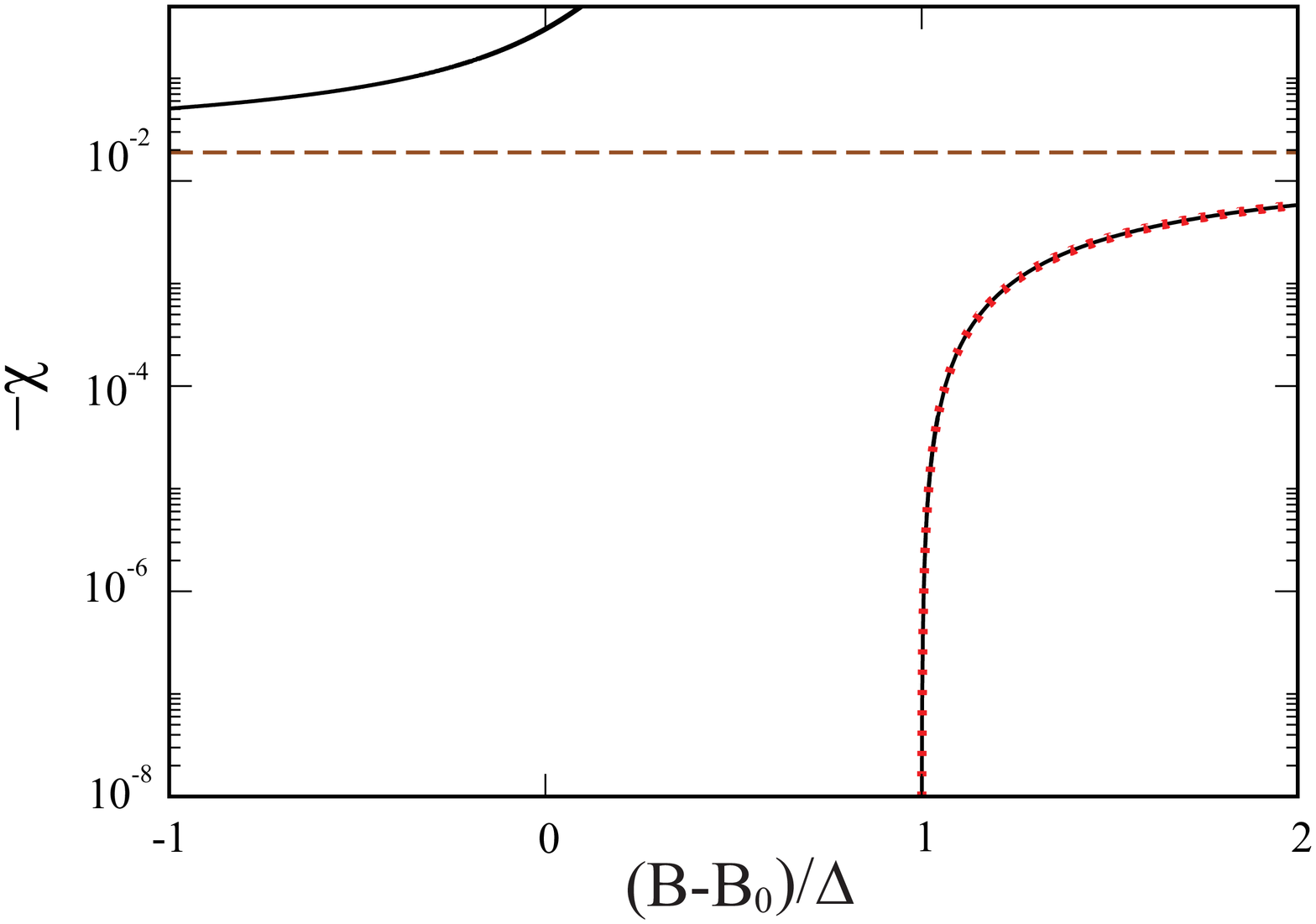} \\
\includegraphics[width=1\textwidth]{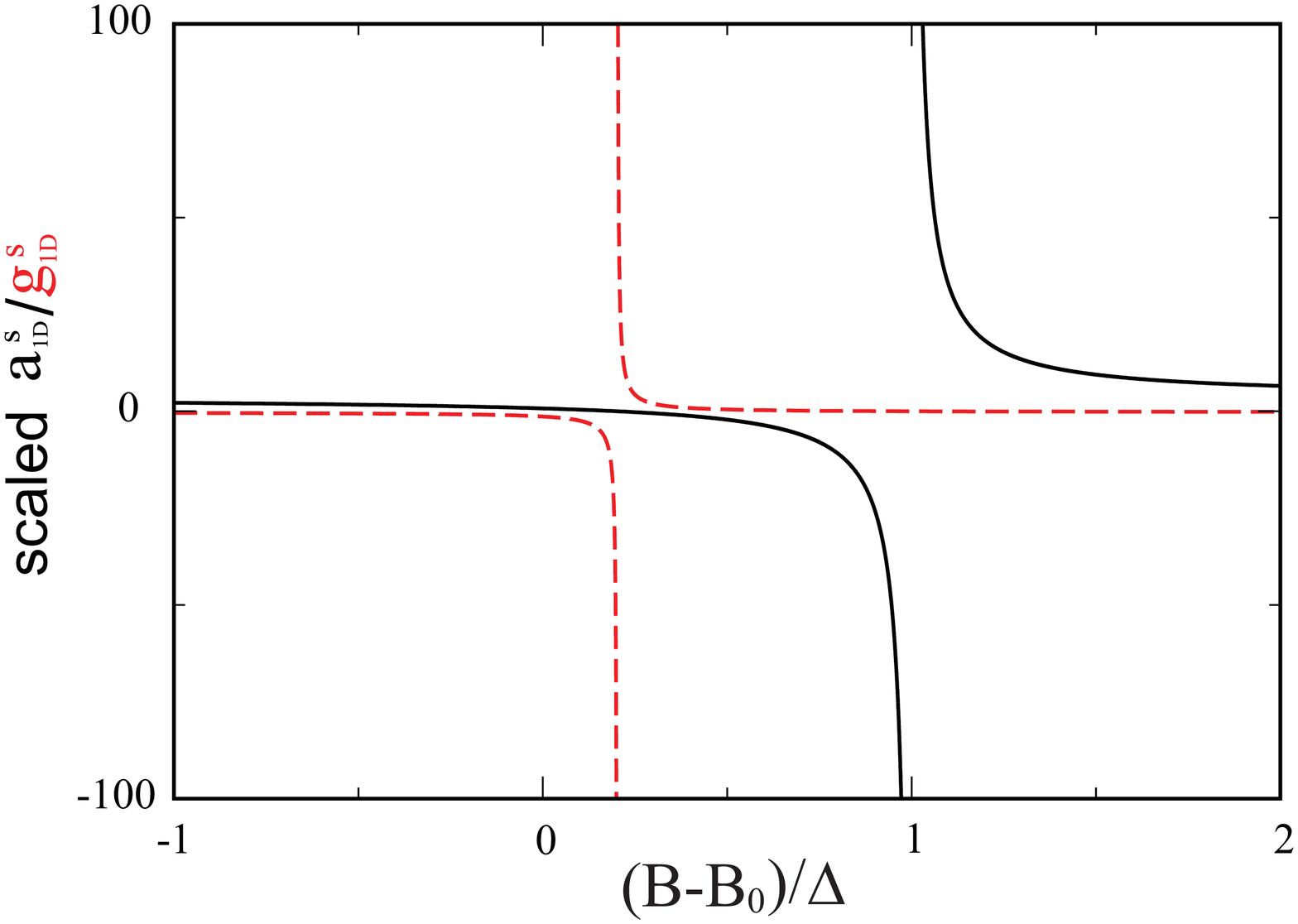}	\\
\end{minipage}
\caption{Bound and scattering properties for two $^6$Li atoms in the waveguide
in the vicinity of the $s$-wave Feshbach resonance at $B_0=832$~G. Upper panel: scaled binding energies calculated
by equations~(\ref{eq_boundlevelinwaveguide2}) (black solid line) and (\ref{eq_boundlevelinwaveguideapproximation})
(red dotted line) are shown. The horizontal brown dashed line corresponds to $-\chi_{\rm bg}=0.0189$.
Lower panel: 1D scattering length $a_{\rm 1D}^s$ scaled by $a_{\perp}$ (black solid line)
and 1D interaction strength $g_{\rm 1D}^s$ scaled by $\hbar^2/\mu{a_{\perp}}$ (red dashed line).
The transverse trapping frequency $\omega_{\perp}$ is set to $2\pi\times14$~kHz in the calculation,
which is realized in \cite{prl:110:203202}. The harmonic length $a_{\perp}$
and van der Waals length $\beta_6$ are 9259 au and 63 au, respectively.}
\label{fig_BoundLevel_a1Dg1D_broadResonance}
\end{figure}
\begin{figure}
\centering
\begin{minipage}[b]{0.5\textwidth}
\includegraphics[width=1\textwidth]{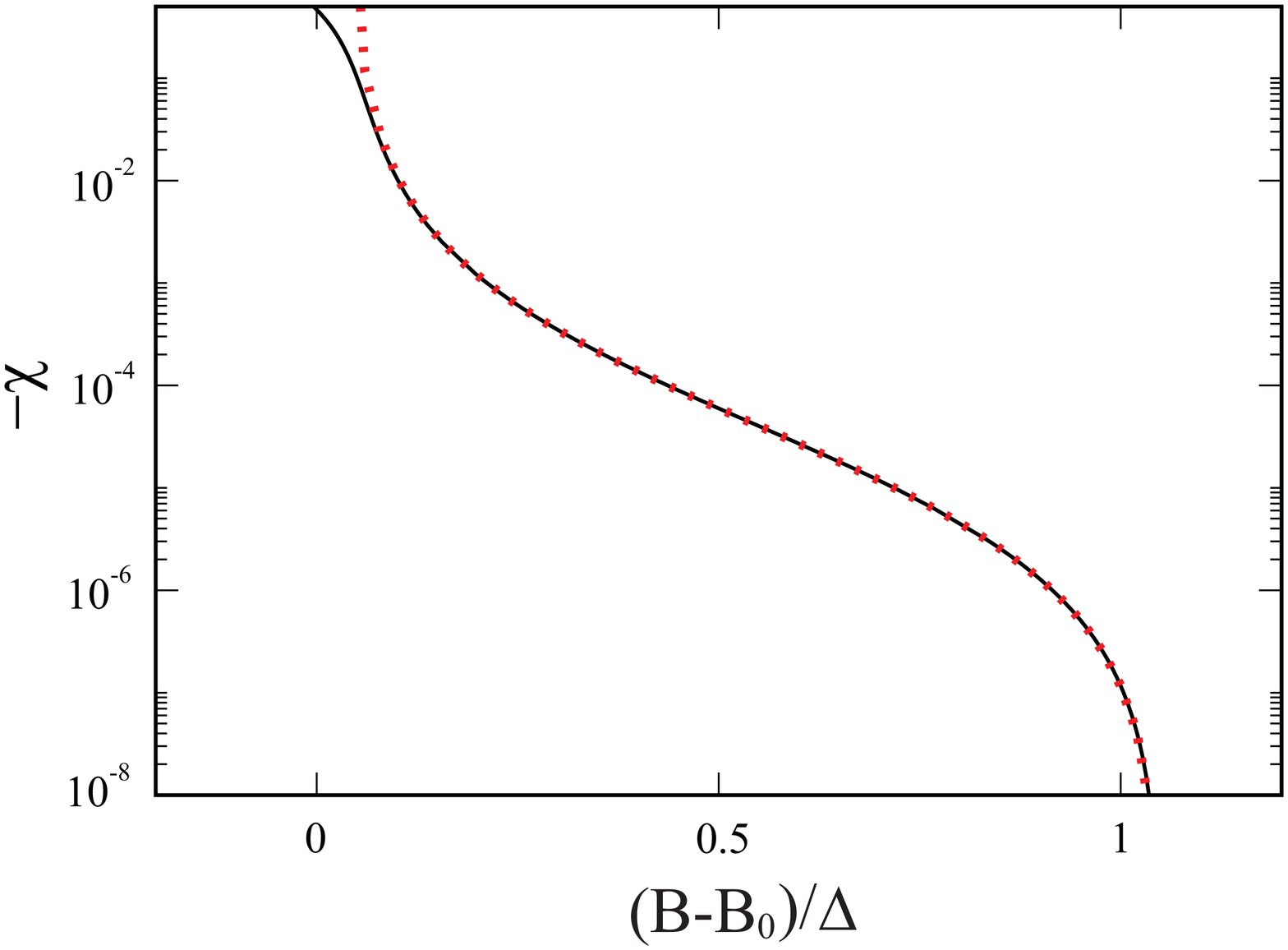} \\
\includegraphics[width=1\textwidth]{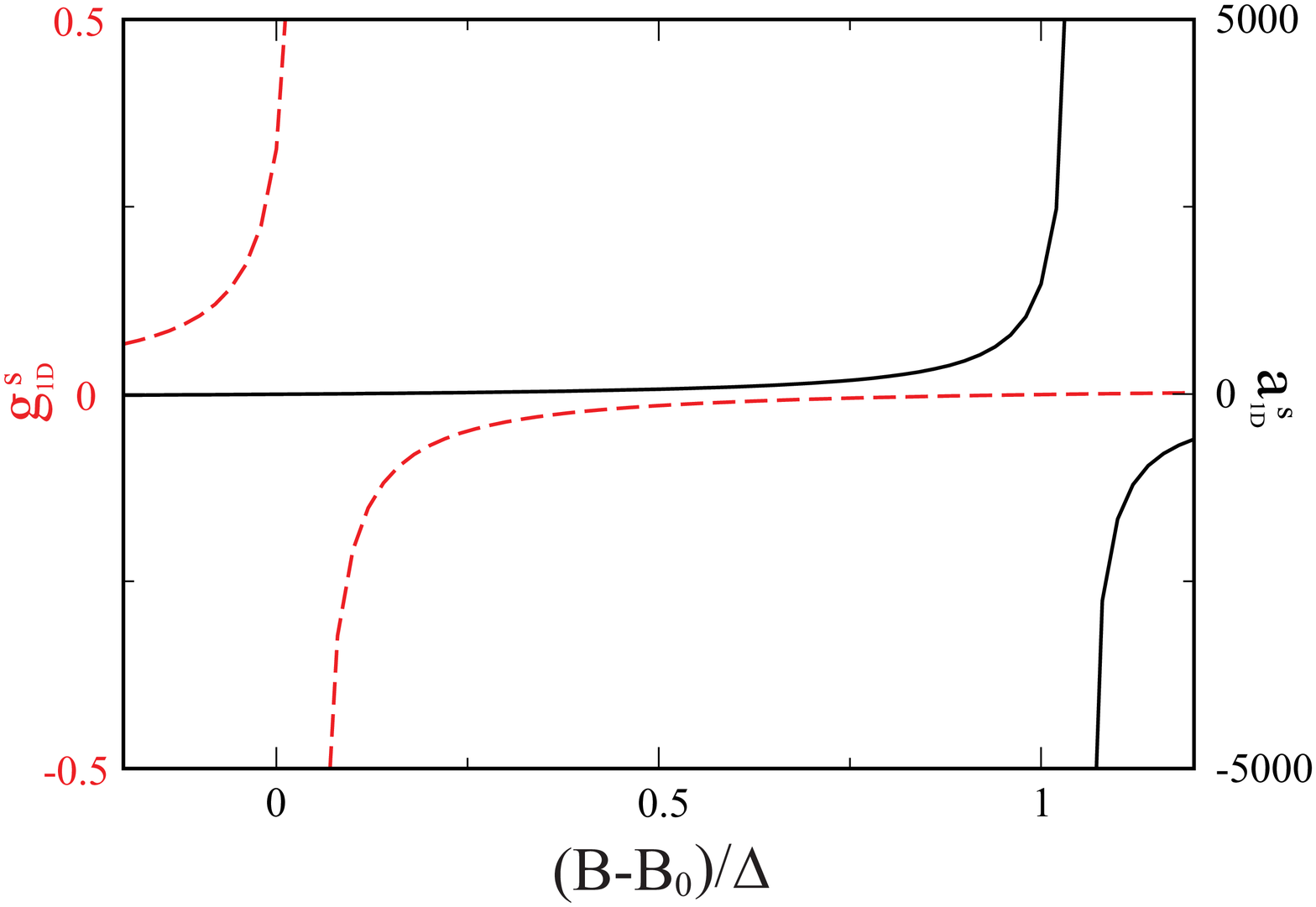}	\\
\end{minipage}
\caption{Bound and scattering properties in a waveguide  
in the vicinity of the $^6$Li $s$-wave Feshbach resonance at $B_0=534.4$~G. Binding energies calculated 
by equations~(\ref{eq_boundlevelinwaveguide2}) (black solid line) and (\ref{eq_boundlevelinwaveguideapproximation})
(red dotted line) are shown in the upper panel. $a_{\rm 1D}^s$ scaled by $a_{\perp}$ (black solid line)
and $g_{\rm 1D}^s$ scaled by $\hbar^2/\mu{a_{\perp}}$ (red dashed line) are displayed in the lower panel. 
The confining frequency is the same as in figure~\ref{fig_BoundLevel_a1Dg1D_broadResonance}.}
\label{fig_BoundLevel_a1Dg1D_narrowResonance}
\end{figure}

\indent Making use of the relation between $K^{\rm 1D}$ and $K^{\rm 3D}$,
and after some algebra \cite{pra:89:052716}, the bound-state equation (\ref{eq_boundlevelinwaveguide})
for $l=0$ can be written as
\begin{equation}
\frac{a_{\perp}}{a_{\rm eff}(E,B)}=-\zeta\left(\frac{1}{2},-\chi\right),
\label{eq_boundlevelinwaveguide2}
\end{equation}
where $a_{\rm eff}(E,B)=-\tan\eta_s(E,B)/k$ is the effective scattering length
introduced in \cite{pra:65:043613, pra:66:013403}, and $\eta_s$ is the $s$-wave scattering phase shift. 
$-\chi=(\hbar\omega_{\perp}-E_{\rm b})/\hbar\omega_{\perp}$ is the scaled binding energy of the bound state 
with energy $E_{\rm b}$, and $\zeta(s,p)$ is the Hurwitz zeta function.
Combining equations~(\ref{eq_K1DPhys}) and (\ref{eq_a1Ds}), the 1D scattering length can be expressed as
\begin{equation}
a_{\rm 1D}^s(B)=-\frac{a_{\perp}}{2}\left[\frac{a_{\perp}}{a_{\rm eff}(E=\hbar\omega_{\perp},B)}+
\zeta\left(\frac{1}{2}\right)\right],
\label{eq_EnergyDependenta1D}
\end{equation}
and $\zeta(s)$ is the Riemann zeta function.
The expressions for the bound state and $a_{\rm 1D}^s$ are the same with that from \cite{JPIV:116:69}
except that the effective scattering length at the zero-point energy of the transverse confinement 
$E=\hbar\omega_{\perp}$ is used instead of the zero-energy scattering length. \\
\indent  The binding energy of two $^6$Li atoms in a waveguide close to the broad resonance I is shown by
the black solid line in the upper panel of figure~\ref{fig_BoundLevel_a1Dg1D_broadResonance}. 
Besides the Feshbach dimer which induces the free space resonance, 
a confinement-induced background dimer emerges in the waveguide due to
the negative free space background scattering length $a_{\rm bg}$ for resonance I \cite{pra:91:042703}. 
Figure~\ref{fig_BoundLevel_a1Dg1D_broadResonance} shows the avoid crossing between the two bound states.
The effective 1D scattering length $a_{\rm 1D}^s$ and
the 1D interaction strength $g_{\rm 1D}^s$ are shown in the lower panel of figure~\ref{fig_BoundLevel_a1Dg1D_broadResonance} 
by the black solid and red dashed lines, respectively.
One can see that the bound state crosses the threshold at the pole of $a_{\rm 1D}^s$ 
when $g_{\rm 1D}^s$ vanishes.  
This agrees with the observation in \cite{pra:86:062713} that the bound state component
in the scattering wavefunction is dramatically enhanced at the zero crossing of the free space 
scattering length $a_{\rm eff}$, not at CIR. It is noted that the zero crossing of $a_{\rm eff}$ 
corresponds to infinite $a_{\rm 1D}^s$ according to equation~(\ref{eq_EnergyDependenta1D}).
In the vicinity of the broad $^6$Li resonance, the association of atom pairs into molecules in a harmonic waveguide 
is expected to happen by sweeping the magnetic field adiabatically across the region $B{\sim}B_0+\Delta$.\\
\indent For a weakly bound state $\chi\rightarrow{0^-}$, $E=2\hbar\omega_{\perp}\left(\frac{1}{2}+\chi\right){\sim}\hbar\omega_{\perp}$,
and the Hurwitz zeta function in equation~(\ref{eq_boundlevelinwaveguide2}) can be approximated by
$\zeta\left(\frac{1}{2},-\chi\right){\sim}\frac{1}{\sqrt{-\chi}}+\zeta\left(\frac{1}{2}\right)$ \cite{handbook}.
Then the binding energy of the weakly bound state can be expressed in terms of $a_{\rm 1D}^s$ as
\begin{equation}
-\chi(B)=\frac{a_{\perp}^2}{4a_{\rm 1D}^s(B)^2}.
\label{eq_boundlevelinwaveguideapproximation}
\end{equation}
The binding energy calculated by equation~(\ref{eq_boundlevelinwaveguideapproximation})
is shown in the upper panel of figure~\ref{fig_BoundLevel_a1Dg1D_broadResonance} by the
red dotted line. \\
\indent Substituting equation~(\ref{eq_sWavePhaseShift})
for $a_{\rm eff}$ in equation~(\ref{eq_EnergyDependenta1D}), one can rewrite the expression
as
\begin{equation}
a_{\rm 1D}^s=a_{\rm bg, 1D}\left(1-\frac{\Delta_{\rm 1D}}{B-B_{\rm 0,1D}}\right),
\label{eq_a1Dsimilartoa3D}
\end{equation}
where $a_{\rm bg,1D}=-\frac{a_{\perp}^2}{2a_{\rm bg}}\left(1+\frac{a_{\rm bg}}{a_{\perp}}\zeta\left(\frac{1}{2}\right)\right)$,
$B_{\rm 0,1D}=B_0+\Delta+\frac{\hbar\omega_{\perp}}{\delta\mu}$, and $\Delta_{\rm 1D}=-\Delta\left(1+\frac{a_{\rm bg}}{a_{\perp}}\zeta\left(\frac{1}{2}\right)\right)^{-1}$.
Equations (\ref{eq_boundlevelinwaveguideapproximation}) and (\ref{eq_a1Dsimilartoa3D}) are similar
as their free space counterparts \cite{rmp:82:1225}, and suggest that a 
Feshbach resonance in 1D \cite{ajp_81_603} is realized effectively with two atoms colliding in
the waveguide around a free space Feshbach resonance \cite{pra:71:012709}. 
The resonant position of the
Feshbach resonance in 1D is $B_{\rm 0,1D}$ which is shifted from the free space resonant
position $B_0$ by $\Delta+\hbar\omega_{\perp}/\delta\mu$. 
It is to be noted that the term $\hbar\omega_{\perp}/\delta\mu$ originates from the energy-dependent term in 
equation~(\ref{eq_sWavePhaseShift}). For broad resonances and close to the threshold, 
the energy-dependence of the phase shift is negligible \cite{rmp:82:1225}, and hence the term $\hbar\omega_{\perp}/\delta\mu$ 
in the expression of $B_{\rm 0,1D}$ can be ignored. This statement is supported by the following quantitative analysis.
Making use of the definition of $R^*$, the term $\hbar\omega_{\perp}/\delta\mu$
can be expressed as $\frac{a_{\rm bg}R^*}{2a_{\perp}^2}\Delta$.
Usually the background scattering length $a_{\rm bg}$ for two atoms colliding in free space
is of the order of the van der Waals length $\beta_6$, which is assumed to be far smaller than $a_{\perp}$.
For broad resonances, $R^*$ is small, and hence $\hbar\omega_{\perp}/\delta\mu$ is negligible compared to $\Delta$.
The 1D resonant position $B_{\rm 0,1D}$ is approximated to be $B_0+\Delta$,
which is independent on the transverse confinement frequency $\omega_{\perp}$. 
This is verified in figure \ref{fig_BoundLevel_a1Dg1D_broadResonance} for
resonance I which shows that $a_{\rm 1D}^s$ diverges at $(B-B_0)/\Delta\sim1$.
At $B=B_{\rm 0, 1D}+\Delta_{\rm 1D}$, $a_{\rm 1D}^s$ equals 
to zero, and $g_{\rm 1D}$ tends to infinity.
The width of the Feshbach resonance in 1D is $\Delta_{\rm 1D}$,
which has the same magnitude as the width $\Delta$ of the free space resonance,
but a different sign.
The background 1D scattering length is $a_{\rm bg, 1D}$,
which is of the order of $a_{\perp}$, and has the opposite sign of $a_{\rm bg}$.
For the $^6$Li resonance considered in figure \ref{fig_BoundLevel_a1Dg1D_broadResonance},
the free space background scattering length $a_{\rm bg}$ is large and negative, i.e.\ we have
$\Delta_{\rm 1D}=-0.8\Delta$, and $a_{\rm bg, 1D}/a_{\perp}=3.6$.
The binding energy of the confinement induced bound state away from
the resonant magnetic field $B_{\rm 0,1D}$ is estimated by substituting 
$a_{\rm bg, 1D}$ into equation~(\ref{eq_boundlevelinwaveguideapproximation}), 
which gives $-\chi_{\rm bg}=\frac{a_{\perp}^2}{4a_{\rm bg, 1D}^2}=0.0189$. 
This value is depicted by the brown dashed line in
the upper panel of figure~\ref{fig_BoundLevel_a1Dg1D_broadResonance}.\\
\indent An analogous investigation has been performed for two fermionic $^6$Li atoms
in a harmonic waveguide around the narrow resonance II, and the
results are shown in figure~\ref{fig_BoundLevel_a1Dg1D_narrowResonance}.
For resonance II, $a_{\rm bg}$ is positive, and $a_{\rm bg,1D}$ is negative.
There is no confinement induced background dimer, and only the Feshbach dimer exists \cite{pra:91:042703}.
For narrow resonances, the phase shift is strongly energy-dependent \cite{rmp:82:1225}, and 
the energy-dependent term in equation~(\ref{eq_sWavePhaseShift}) cannot be neglected. 
Accordingly, the term $\hbar\omega_{\perp}/\delta\mu$
in the expression of $B_{\rm 0, 1D}$ cannot be neglected. 
In figure~\ref{fig_BoundLevel_a1Dg1D_narrowResonance}, it is clearly shown that $a_{\rm 1D}^s$ diverges
at a magnetic field strength different from $(B-B_0)/\Delta=1$.\\
\indent In free space, a weakly bound state exists only on the positive side
of the free space scattering length close to resonance \cite{rmp:82:1225}.
In the waveguide, it has been verified experimentally that
the weakly bound state exists for both positive and negative free space scattering length \cite{prl:94:210401}.
Figures~\ref{fig_BoundLevel_a1Dg1D_broadResonance} and \ref{fig_BoundLevel_a1Dg1D_narrowResonance}
show that, in terms of the effective one-dimensional scattering length $a_{\rm 1D}^s$,
similar statements like those in free space can be made in the presence of the waveguide, i.e.\
the weakly bound state exists only on the positive side of $a_{\rm 1D}^s$.
\subsection{$p$-wave case}
\begin{figure}
\centering
\begin{minipage}[b]{0.5\textwidth}
\includegraphics[width=1\textwidth]{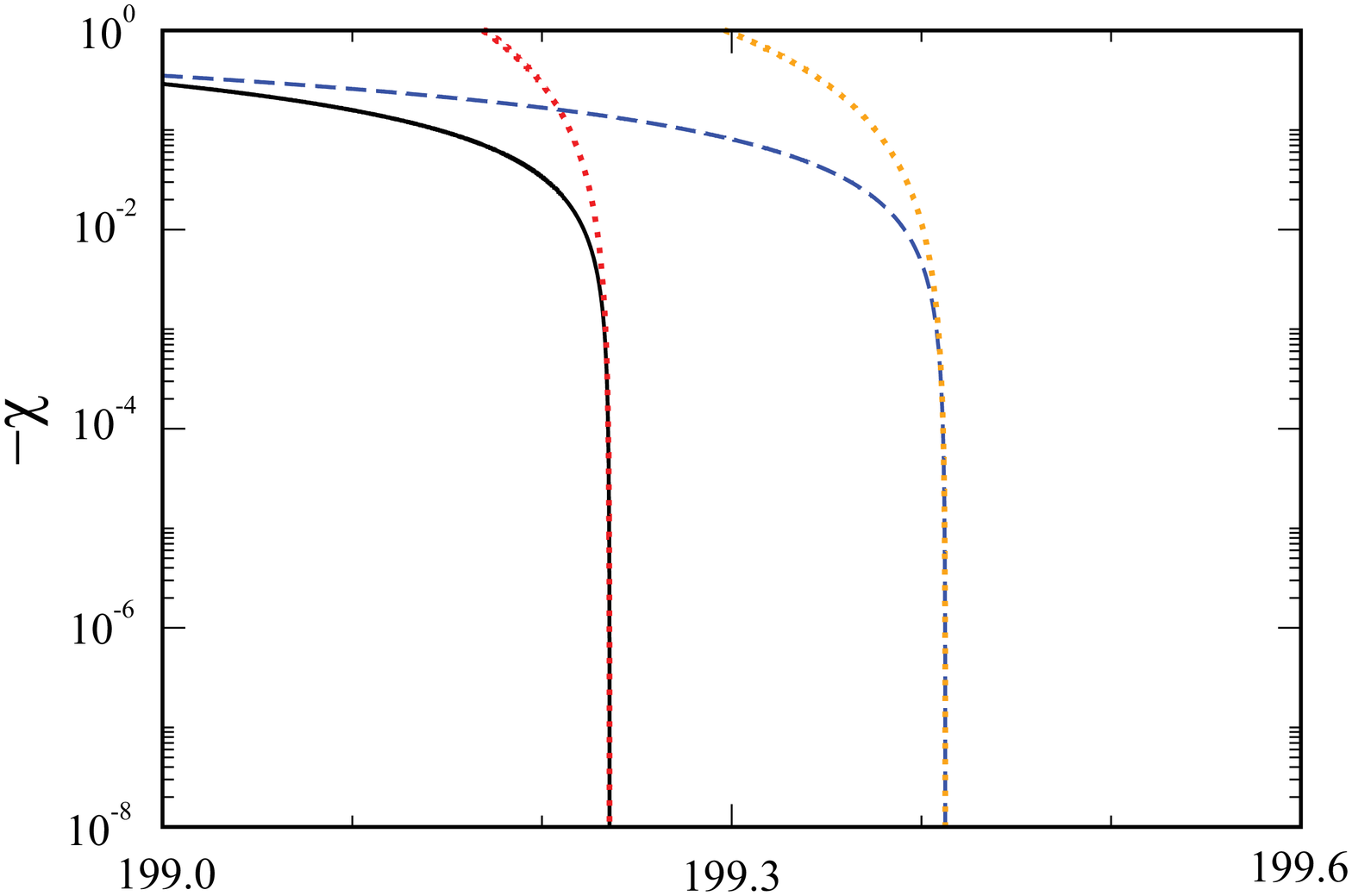} \\
\includegraphics[width=1\textwidth]{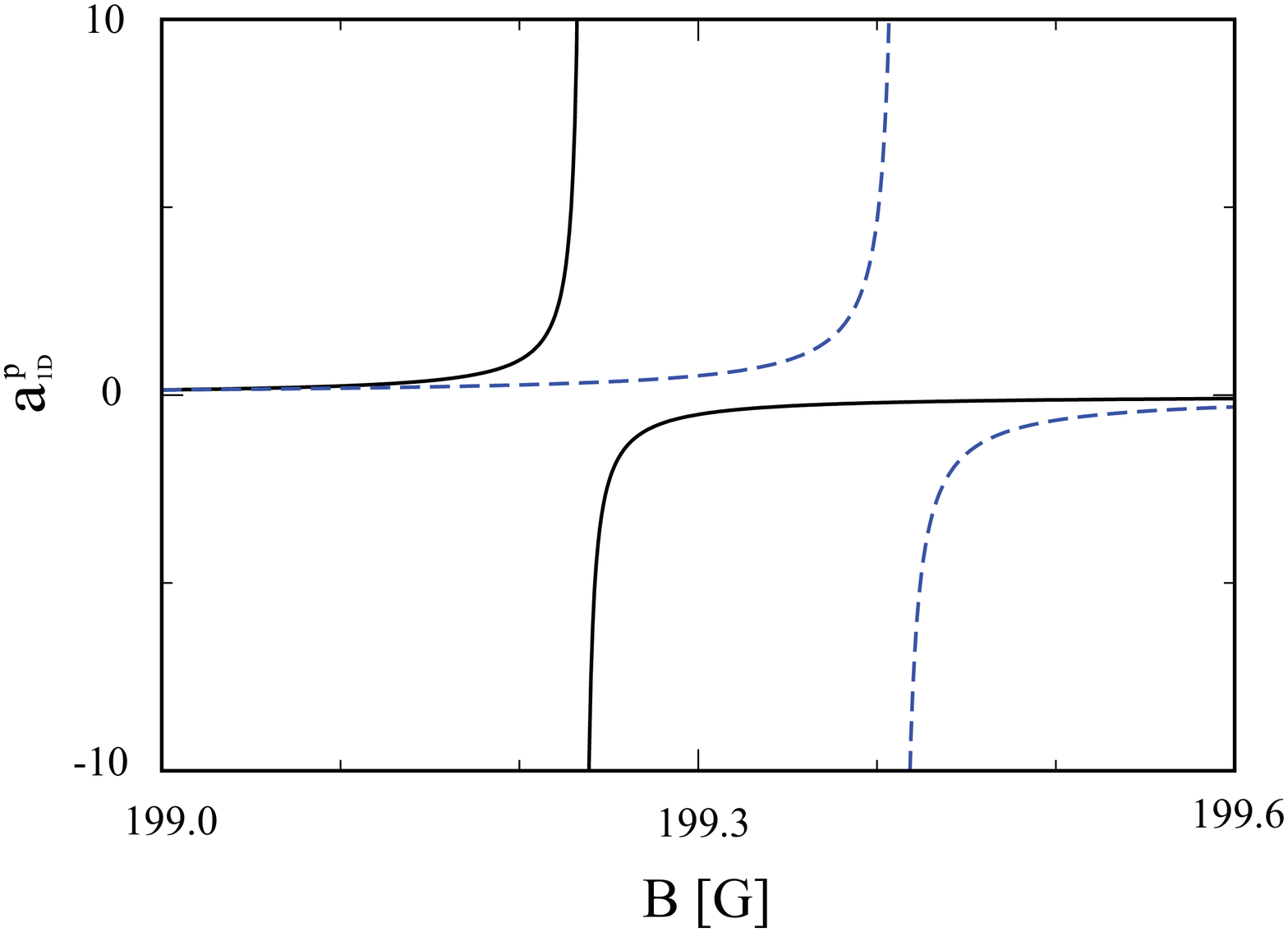}	\\
\end{minipage}
\caption{Binding energy (upper panel) and effective 1D scattering length (lower panel) 
in a harmonic waveguide close to the $^{40}$K $p$-wave Feshbach resonance \cite{pra:69:042712}. 
Results with confining frequency $\omega_{\perp}=2\pi\times$69 and 100~kHz are shown 
in black solid and blue dashed lines, respectively. The red and orange dotted lines in the upper 
panel are the binding energies calculated by equation~(\ref{eq_boundlevelinwaveguideapproximation}). 
The 1D scattering length $a_{\rm 1D}^p$ shown in the lower panel is scaled by $a_{\perp}$.
}
\label{fig_BoundLevel_a1Dg1D_pWaveResonance}
\end{figure}
\indent Let us now focus on the $p$-wave interaction in the waveguide.
Substituting $K^{\rm 1D}$ by $UK^{\rm 3D}U^T$ and following the derivation in \cite{prl:92:133202,pra:89:052716}, 
the bound-state equation (\ref{eq_boundlevelinwaveguide}) for $l=1$ becomes
\begin{equation}
12\zeta\left(-\frac{1}{2},-\chi\right)=\frac{a_{\perp}^3}{V_p(E,B)},
\label{eq_boundlevelinwaveguidePWave}
\end{equation}
where $V_p(E,B)=-\tan\eta_p(E,B)/k^3$, and $\eta_p$ is the $p$-wave scattering phase shift. 
According to equations~(\ref{eq_K1DPhys}) and (\ref{eq_a1Dp}), the effective 1D scattering length 
$a_{\rm 1D}^p$ is 
\begin{equation}
a_{\rm 1D}^p(B)=6a_{\perp}\left[\frac{a_{\perp}^3}{V_p(E=\hbar\omega_{\perp},B)}-12\zeta\left(-\frac{1}{2}\right)\right]^{-1},
\end{equation}
which reproduces the corresponding expressions in \cite{prl:92:133202,prl:100:170404}. 
For a weakly bound state $\chi\rightarrow{0^-}$, the Hurwitz zeta function 
in equation~(\ref{eq_boundlevelinwaveguidePWave}) can be approximated by 
$\zeta\left(-\frac{1}{2},-\chi\right){\sim}\sqrt{-\chi}+\zeta\left(-\frac{1}{2}\right)$ 
\cite{handbook}. The relation between the binding energy of the weakly bound state and 
1D scattering length, equation~(\ref{eq_boundlevelinwaveguideapproximation}), also 
applies for the $p$-wave case if $a_{\rm 1D}^s$ is substituted by $a_{\rm 1D}^p$. \\
\indent The binding energy and effective 1D scattering length for two $^{40}$K atoms 
in a harmonic waveguide interacting in the vicinity of a $p$-wave Feshbach resonance \cite{pra:69:042712} 
are shown in figure~\ref{fig_BoundLevel_a1Dg1D_pWaveResonance}. 
Calculations are perfromed with two confining frequencies $\omega_{\perp}=2\pi\times$69~kHz (black solid line) 
and $2\pi\times$100~kHz (blue dashed line). 
The binding energy determined via equation~(\ref{eq_boundlevelinwaveguideapproximation}),
which is valid for a weakly bound state, is depicted in red and orange dotted lines in the upper panel. 
The results show that the confining frequency 
can be used to tune the bound state and scattering properties. \\
\indent As in the $s$-wave case, 
the bound state in the waveguide crosses the scattering threshold at the magnetic field 
where $a_{\rm 1D}^p$ diverges. For $p$-wave scattering, $g_{\rm 1D}^p$ is proportional to $a_{\rm 1D}^p$,   
and a $p$-wave CIR occurs at the magnetic field strength where $a_{\rm 1D}^p$ is divergent. 
Hence, the adiabatic molecular formation happens at the position of the CIR for $p$-wave interaction \cite{prl:95:230401}, 
in contrast to the $s$-wave case where the molecular formation occurs at $g_{\rm 1D}^s=0$ other than at CIR.\\
\section{Conclusion} 
\indent The bound state and two-body collisions in a harmonic waveguide close to 
free space Feshbach resonances have been investigated 
by using the local frame transformation approach. 
As an extension to previous studies \cite{prl:92:133202,pra:89:052716,pra:92:022706}, 
the energy and magnetic field dependent free space phase shifts due to a 
realistic multichannel interatomic interaction are adopted in the LFT method.  
The LFT method, which relies on the length scale separation $\beta_6\ll\omega_{\perp}$, 
complements the zero-range \cite{pra:72:053625, pra:73:052709, pra:86:033601, pra:91:043622} and 
finite-range \cite{pra:91:042703} two-channel models which have been used to explore 
the two-body properties in the waveguide close to Feshbach resonances. \\
\indent The position of the association of atoms into molecules in a harmonic waveguide 
during an adiabatic sweep of the magnetic field has been investigated, 
and its relation with the position of the confinement induced resonance, 
indicated by an infinite one-dimensional interaction strength $g_{\rm 1D}$, has been discussed. 
For both $s$ and $p$ wave interaction, 
the least bound state in the waveguide 
crosses the scattering threshold at the magnetic field $B_{\rm 0, 1D}$ 
where the effective one-dimensional scattering 
length $a_{\rm 1D}$ diverges. For the $s$-wave case, $a_{\rm 1D}$ and $g_{\rm 1D}$ 
are inversely proportional to each other, and the position of molecular formation 
differs from the position of the $s$-wave CIR.
For $p$-wave interaction, $a_{\rm 1D}$ is proportional to $g_{\rm 1D}$, and 
the position of the molecular formation coincides with the position of the $p$-wave CIR. 
Moreover, in the vicinity of broad $s$-wave 
Feshbach resonances, the magnetic field $B_{\rm 0, 1D}$ is nearly independent 
on the transverse confining frequency $\omega_{\perp}$, and is approximately $B_0+\Delta$, 
where $B_0$ and $\Delta$ are the free space resonance position and width. 
Close to $p$-wave and narrow $s$-wave Feshbach resonances, the free space phase shift is strongly energy-dependent.  
The zero-point energy $\hbar\omega_{\perp}$ of the transverse confinement plays a role in determining the value of $B_{\rm 0, 1D}$, 
and $B_{\rm 0, 1D}$ changes as $\omega_{\perp}$ varies.
\section*{Acknowledgments}
The authors thank B. Heß, V. S. Melezhik, and C. H. Greene for fruitful discussions.
Comments on the manuscript by V. S. Melezhik are appreciated.
G. W. acknowledges a fellowship from the Alexander von Humboldt Foundation. 
P. G. acknowledges financial support by the NSF through grant PHY-1306905.\\



\end{document}